\documentclass[runningheads]{llncs}
\usepackage{graphicx}
\usepackage{hyperref}
\usepackage{url}
\usepackage{soul}
\usepackage{subcaption}
\usepackage{caption}

\begin{document}

\title{Family-Based Fingerprint Analysis:\\ A Position Paper}
\titlerunning{Family-Based Fingerprint Analysis}
\author{
Carlos Diego N. Damasceno\inst{1}\orcidID{0000-0001-8492-7484}
\and Daniel Strüber\inst{1,2}\orcidID{0000-0002-5969-3521}
}
\authorrunning{Damasceno and Strüber}
\institute{Radboud University, Nijmegen, NL \\ 
\email{d.damasceno@cs.ru.nl} 
\and
Chalmers University of Technology, Gothenburg, SE\\
\email{danstru@chalmers.se}
}
\maketitle              %
\begin{abstract}
Thousands of vulnerabilities are reported on a monthly basis to security repositories, such as the National Vulnerability Database. Among these vulnerabilities, software misconfiguration is one of the top 10 security risks for web applications. With this large influx of vulnerability reports, software fingerprinting has become a highly desired capability to discover distinctive and efficient signatures and recognize reportedly vulnerable software implementations. Due to the exponential worst-case complexity of fingerprint matching, designing more efficient methods for fingerprinting becomes highly desirable, especially for variability-intensive systems where optional features add another exponential factor to its analysis. This position paper presents our vision of a framework that lifts model learning and family-based analysis principles to software fingerprinting. In this framework, we propose unifying databases of signatures into a featured finite state machine and using presence conditions to specify whether and in which circumstances a given input-output trace is observed. We believe feature-based signatures can aid performance improvements by reducing the size of fingerprints under analysis.
\end{abstract}
\section{Introduction}

Automatically recognizing vulnerable black-box components is a critical requirement in security analysis, especially considering the fact that modern systems typically include components borrowed from free and open-source projects. Besides, with the large influx of versions released over time and vulnerabilities reported in security data sources, such as the National Vulnerability Database (NVD) \cite{nvd_national_2022}, engineers should dedicate a special attention to the efficiency and the scalability of techniques for automated software analysis. Providing such a capability can dramatically reduce engineer's workload and greatly increase the efficiency as well as the accuracy of security analysis. One of such techniques is software fingerprinting \cite{shu_formal_2011}.

Software fingerprinting aims to produce a distinctive and efficient \textit{signature} from syntactic, semantic, or structural characteristics of a system under test (SUT). It is an important technique with many security applications, ranging from malware detection, digital forensics, copyright infringement, to vulnerability analysis \cite{alrabaee_survey_2022}. To produce a signature that is both expressive and identifiable, fingerprint discovery and matching can be pursued using different techniques \cite{alrabaee_survey_2022}, such as text-based models (e.g., code instruction or strings), structural models (e.g., call graphs or control/data-flow graphs), and behavioral-based models (e.g., execution traces and finite state machines).
When source code is unavailable, model learning \cite{angluin_learning_1987,vaandrager_model_2017} and testing \cite{broy_model-based_2005} techniques may be used as means to capture the behavioral signatures of an SUT in terms of states and transitions of a finite state machine. 

Model learning has emerged as an effective bug-finding technique for black-box hardware and software components \cite{vaandrager_model_2017}. In active model learning \cite{angluin_learning_1987}, a learning algorithm interacts with an SUT to construct and validate a hypothesis $\mathcal{H}$ about the “language” of its external behavior. In general, this hypothesis is expressed as a Mealy finite state machine (FSM) that, once established, can be deployed as a \textit{behavioral signature} to recognize an SUT. 
Model learning has been reported effective in building models of different network protocols, including TCP \cite{fiterau-brostean_combining_2016}, TLS \cite{de_ruiter_tale_2016,janssen_fingerprinting_2021}, Bluetooth \cite{pferscher_fingerprinting_2021}, and MQTT \cite{tappler_model-based_2017}.

Once a group of fingerprints is produced for a set of SUTs, two naive approaches may take place to identify whether an unidentified SUT matches with any of the signatures in a group of fingerprints \cite{janssen_fingerprinting_2021}: (a) re-run model learning over the unidentified SUT and compare the resulting hypothesis to all known signatures; (b) perform conformance testing \cite{broy_model-based_2005} for each model to see which one matches. While both methods can be effective, they are resource and time intensive and hence, inefficient for large groups of candidate fingerprints. As a matter of fact, active group fingerprinting has an exponential worst-case complexity for the number of fingerprints in a database of signatures \cite{shu_formal_2011}. Therefore, finding more efficient ways to perform fingerprint group matching becomes highly desirable.

Fingerprinting is especially challenging in variability-intensive systems, in which particular system variants can be generated by switching features on or off. 
Optional features lead a combinatorial explosion of the overall set of possible system variants and hence, significant challenges for software analyses \cite{elmaghbub_lora_2021}. A recent survey indicates that  software security of variability-intensive systems is an under-studied topic \cite{kenner_safety_2021}. 
To the best of our knowledge, fingerprinting in particular has not been addressed in this context. To date, Shu et al. \cite{shu_formal_2011} and Janssen \cite{janssen_fingerprinting_2021} are the most prominent studies exploring model learning \cite{van_den_bos_state_2021} and conformance testing \cite{lee_principles_1996} in fingerprint group matching. Nevertheless, further investigations are still needed to evaluate the efficiency and scalability of their approaches in large fingerprint databases \cite{janssen_fingerprinting_2021}, as expected in variability-intensive systems. In this paper, we envision optimizing fingerprinting techniques towards variability-intensive and evolving systems.

In our vision, we propose principles from variability-aware software analysis \cite{thum_classification_2014} as means to achieve an efficient framework for family-based fingerprint discovery and matching.
The term \textit{family-based}  \cite{thum_classification_2014} refers to an analysis that is performed at the level of the whole product line, instead of individual products,  thus allowing to efficiently derive statements about \textit{all}  products.
In our proposed framework, we combine groups of behavioral signatures into a family model, e.g., featured finite state machine \cite{fragal_extending_2018}, and use presence conditions to specify whether (and in which circumstances) a given input-output (IO) trace can be observed. In combination with SAT/SMT solvers and state-based model comparison algorithms \cite{walkinshaw_automated_2013,damasceno_learning_2021}, this family-based representation can pave the way for efficient fingerprint discovery and matching techniques where the size of the fingerprints under analysis can be reduced in orders of magnitude.
It would also contribute to addressing the general lack of family-based analyses in the field: Kenner et al's survey \cite{kenner_safety_2021} mentions a single previous family-based security analysis \cite{peldszus2018model}.

This position paper is organized as follows: In section \ref{sec:fgpt}, we introduce software fingerprinting, with an emphasis in active model learning \cite{vaandrager_model_2017}. In section \ref{sec:fb_fgpt}, we draw our vision of family-based fingerprint analysis upon the concept of family-based analysis \cite{thum_classification_2014} and testing \cite{fragal_extending_2018}. We close this article, in section \ref{sec:concl}, with our final remarks about this framework for family-based fingerprint analysis.

\section{Software Fingerprinting}
\label{sec:fgpt}

Software fingerprinting aims at discovering a distinctive and efficient signature of syntactic, semantic, or structural characteristics of a SUT and matching unidentified SUTs against one or more fingerprints in a database. It is a fundamental approach with various applications in software security, including malware detection, software infringement, vulnerability analysis, and digital forensics \cite{alrabaee_survey_2022}. To construct signatures that are both expressive and identifiable, fingerprint discovery and matching can be addressed using different kinds of techniques. In this work, we focus on active model learning \cite{angluin_learning_1987} as a means to achieve fingerprint discovery and matching \cite{shu_formal_2011,janssen_fingerprinting_2021}. 

\subsection{Model Learning}
\begin{figure}[!ht]
    \centering
    \includegraphics[width=0.8\textwidth]{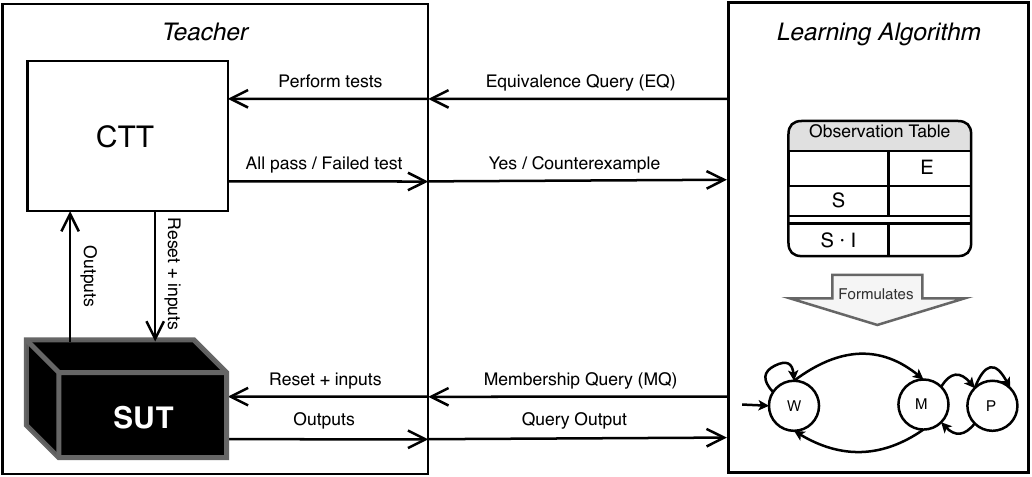}
    \vspace{-0.2cm}
    \caption{The MAT framework (adapted from \cite{vaandrager_model_2017})}
    \vspace{-0.5cm}
    \label{fig:matFramework}
\end{figure}

Active model learning \cite{angluin_learning_1987} has been proven effective in fingerprinting behavioral signatures from black-box software implementations \cite{fiterau-brostean_analysis_2020,janssen_fingerprinting_2021,shu_formal_2011,de_ruiter_tale_2016}. 
For an overview on model learning, we refer the interested reader to Frits Vaandrager's cover article\footnote{In fact, we would like to thank for this well-crafted introduction that sparked our interest to the topic and led to the initial ideas of the first author's doctoral thesis.} of the Communications of the ACM Volume 60 \cite{vaandrager_model_2017}.
Active model learning is often described in terms of the Minimally Adequate Teacher (MAT) framework \cite{angluin_learning_1987} shown in Fig.~\ref{fig:matFramework}.

In the MAT framework, a learning algorithm is used to interact with a black-box system and construct a hypothesis $\mathcal{H}$ about the “language” of a system's external behavior. To construct $\mathcal{H}$, the learning algorithm poses membership queries (\texttt{MQ}) formed by prefixes and suffixes to respectively access and distinguish states in the SUT. Traditionally, these input sequences are maintained in an observation table that guides the formulation of a hypothesis $\mathcal{H}$ of the SUT behavior as a finite state machine (FSM) \cite{broy_model-based_2005}. 

Once a hypothesis is formulated, equivalence queries (\texttt{EQ}) are used to check whether $\mathcal{H}$ fits in the SUT behavior, otherwise it replies a counterexample that exposes any differences. \texttt{EQ}s are typically derived using conformance testing techniques \cite{broy_model-based_2005}. To handle more complex behavior, learning algorithms can also enrich hypotheses with time intervals \cite{tappler_time_2019,aichernig_passive_2020} and data guards \cite{shu_formal_2011}. Whenever a hypothesis is consistent with an SUT, it can be deployed as a fingerprint  \cite{vaandrager_model_2017,janssen_fingerprinting_2021,alrabaee_survey_2022}. 

\subsection{A methodology and taxonomy for formal fingerprint analysis}

Software fingerprinting has been the focus of previous research from multiple angles \cite{alrabaee_survey_2022}. A formal methodology for fingerprinting problems is introduced by Shu et al. \cite{shu_formal_2011}. They introduce the Parameterized Extended Finite State Machine (PEFSM) model as an extension of the FSM formalism that incorporates state variables, guards, and parameterized IO symbols to represent behavioral signatures of network protocols. Using the PEFSM model, the authors discuss a taxonomy of network fingerprinting problems where these are distinguished by their type (active or passive experiments), and goal (matching or discovery). A summary of the taxonomy for fingerprinting problems is shown in Table \ref{tab:fgproblems}.

\begin{table}[!ht]
\centering
\vspace{-0.3cm}
\resizebox{0.9\linewidth}{!}
{%
\begin{tabular}{|c|l|l|}
\hline
Fingerprinting          & \multicolumn{2}{c|}{Experiment type}                            \\ \cline{2-3}
problem                 & \multicolumn{1}{c|}{Active}      & \multicolumn{1}{c|}{Passive} \\ \hline
Single matching         & Conformance testing              & Passive testing              \\ \hline
Group matching          & Online matching separation       & Concurrent passive testing   \\ \hline
Discovery with spec.    & Model enumeration and separation & Back-tracking based testing  \\ \hline
Discovery without spec. & Model learning                   & No efficient solution        \\ \hline
\end{tabular}
}
\caption{Taxonomy of fingerprinting problems (adapted from \cite{shu_formal_2011})}
\vspace{-0.7cm}
\label{tab:fgproblems}
\end{table}

In active fingerprinting, security analysts are able to pose queries to an unidentified SUT whenever they want. In contrast, in passive experiments, fingerprint analysis is limited to a finite set of IO traces as source of information. While active experiments are known to be more effective for providing freedom to query as much as wanted, passive experiments have the advantage that the SUT stays completely unaware that it is under analysis. The process of building a fingerprint signature for an SUT is named \textit{fingerprint discovery}.

In fingerprint discovery \cite{shu_formal_2011}, the goal is to systematically build a distinctive and efficient fingerprint for a SUT. This can be performed by retrieving as much information as possible with the guidance of a pre-existing specification. Otherwise, if no specification is available, model learning \cite{vaandrager_model_2017} can be still applied to build behavioral signatures. Once a database of signatures is established, the task of \textit{fingerprint matching} can take place. 

Typically, the goal of fingerprint matching is to determine whether the behavior of an unidentified SUT matches a single fingerprint signature. However, in cases where there are multiple signatures, it may be interesting to consider matching the SUT against a set of fingerprints of different versions of an implementation \cite{shu_formal_2011}.

\textit{Active group fingerprinting} has been reported to require an exponential worst-case execution time defined by the number of fingerprints in a group \cite{shu_formal_2011}. Therefore, it is highly desirable to have group matching approaches that are more efficient than checking fingerprints one by one.

\begin{example}{\textit{(Running example of fingerprint analysis)}}
In Fig.~\ref{fig:three_fsms}, we depict three alternative versions of an FSM describing the behavior of characters in a game platform, namely $\mathtt{v1,v2,v3}$. 

\begin{figure}
    \centering
    \vspace{-0.4cm}
    \includegraphics[width=0.45\textwidth]{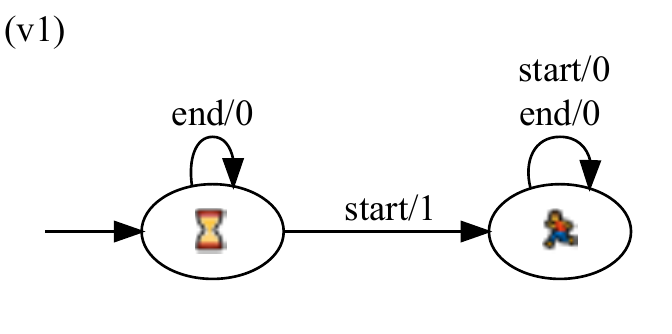}
    \includegraphics[width=0.45\textwidth]{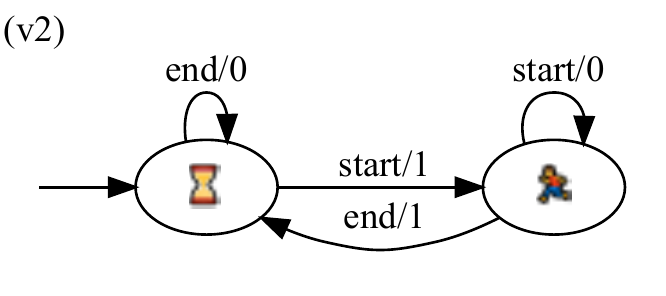}
    
    \vspace{-0.3cm}
    \includegraphics[width=0.6\textwidth]{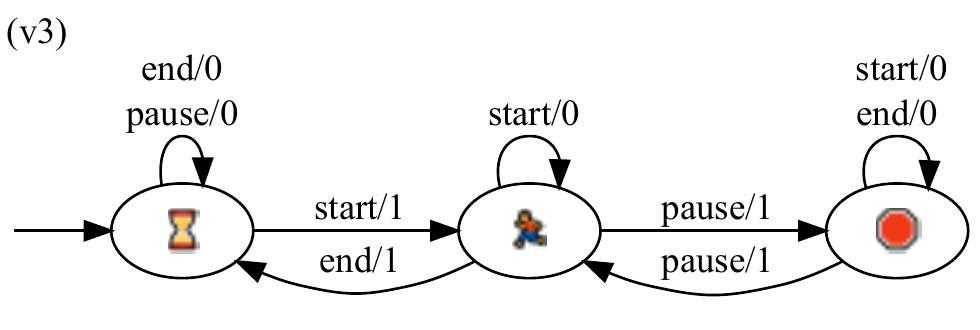}
    \caption{Family of product FSMs}
    \label{fig:three_fsms}
\end{figure}

In the first version \texttt{v1}, we have a character that stays in constant movement, once it starts walking. In version \texttt{v2}, the character can toggle its moving mode. And, in version \texttt{v3}, the character skills are extended with another feature to temporary \texttt{pause} its movement. To distinguish versions \texttt{v1} and \texttt{v2}, we have the input sequence $\mathtt{start\cdot end}$.
\end{example}

\subsubsection{Limitations and Related Work}

The algorithms for fingerprint matching introduced by Shu et al. \cite{shu_formal_2011} have been specifically designed for PEFSMs. Hence, they cannot be directly applied to other notations, such as Mealy machines \cite{vaandrager_model_2017,vaandrager_new_2022} and timed automata \cite{tappler_time_2019,aichernig_passive_2020}; that have more consolidated and ongoing research. To fill this gap, Janssen \cite{janssen_fingerprinting_2021} introduced two novel methods for group fingerprinting matching in his Master's dissertation, under the supervision of prof. Frits Vaandrager. 

In this work, Janssen \cite{janssen_fingerprinting_2021} explores state-of-the-art conformance testing techniques \cite{van_den_bos_state_2021} in active fingerprint group matching. Despite the empirical evidences using an extensive list of TLS implementations, the author points out that further research is still needed to evaluate the efficiency and scalability of their fingerprint matching methods when models are added over time \cite{janssen_fingerprinting_2021}. This limitation becomes particularly interesting if we consider the large number of release versions that can emerge over time and the influx of vulnerability reports available in security databases. For instance, at the moment this manuscript was produced, the GitHub repository of the OpenSSL project \cite{openssl_foundation_inc_openssl_2022} has 338 release versions and more than 31 thousand commits and, the NVD has more than 300 vulnerabilities associated with the keyword ``\textit{openssl}''.
This reinforces the need for designing fingerprinting techniques able to efficiently handle large sets of signatures.

\section{Family-Based Fingerprint Analysis}
\label{sec:fb_fgpt}

As previously discussed, the efficiency of fingerprinting heavily depends on the number of fingerprints under analysis. In fact, the size of a candidate group of fingerprints is an exponential factor in the worst-case complexity of fingerprint group matching \cite{shu_formal_2011}. In variability-intensive systems, this factor may become more noticeable because the number of valid products is up-to exponential in the number of features \cite{thum_classification_2014}. Thus, to minimize costs and effort, while maximizing the effectiveness, we propose looking at fingerprint discovery and matching from a feature-oriented perspective \cite{kang_feature-oriented_1990}. 

Feature modeling allows software engineers to design and manage families of similar, yet customized products by enabling or disabling features. A feature is any prominent or distinctive user-visible behavior or characteristic of a software system \cite{kang_feature-oriented_1990}. Features are typically managed in association with other assets, including feature models \cite{kang_feature-oriented_1990}, source code \cite{apel_feature-oriented_2013}, and test models \cite{fragal_extending_2018}. 

In fingerprinting, the notion of features may be used to capture variability in IO interfaces, optional build parameters, or even release version identifiers. However, when fingerprinting variability-intensive, evolving software systems, it becomes essential to represent behavioral signatures in a way that is succinct \cite{classen_featured_2013,fragal_validated_2017} and aid the design and implementation of  \textit{variability-aware} analysis strategies \cite{thum_classification_2014}. To pursue performance improvements, there is a research direction dedicated to raise variability-awareness in software analysis by lifting modeling languages and analysis strategies to the so called family-based level \cite{thum_classification_2014}.

\subsection{Family-Based Modeling and Analysis}

In family-based analysis, domain artifacts, such as feature models \cite{kang_feature-oriented_1990}, are exploited to efficiently reason about product variants and feature combinations. To make it feasible, software modeling and analysis principles are extended to become aware of variability knowledge and avoid redundant computations across multiple products; an issue that typically occurs when standard software analysis is applied in an exhaustive, product-based fashion \cite{thum_classification_2014}. 

Product-based analysis techniques are known to be effective but \textit{infeasible} because of the potentially exponential number of valid implementations; or, in the best case, \textit{inefficient}, due to redundant computations over assets shared among multiple products \cite{thum_classification_2014}.

Family-based analysis operates on a unified representation of a family of product-specific representations, namely the \textit{family model}. A Featured Finite State Machine (FFSM) \cite{fragal_extending_2018} is one example of variability-aware modeling notation proposed to express families of FSMs as a unified artifact. In FFSMs, states and transitions are annotated with presence conditions described as propositional logic formulae defined over the set of features. These FSM fragments are called conditional transition \cite{fragal_extending_2018} as they occur only when the feature constraints involved in a concerned state or transition are satisfied.

Using SAT solvers, family models are amenable to automated derivation of product-specific models \cite{fragal_validated_2017}, family-based model checking \cite{classen_featured_2013}, and configurable test case generation \cite{fragal_extending_2018}, where redundant analysis over shared states/transitions are mitigated. Thus, the cost of family-based analysis becomes determined by the feature size and amount of feature sharing, instead of the number of valid products \cite{thum_classification_2014}. 

To guide the creation and maintenance of family models, recent studies have proposed the application of model comparison algorithms, such as LTS\_diff \cite{walkinshaw_automated_2013} and FFSM\_diff \cite{damasceno_learning_2021}, to match and merge product-specific FSMs. These approaches can provide efficient means to find differences between models \cite{walkinshaw_automated_2013} and produce succinct FFSM representations from families of FSMs \cite{damasceno_learning_2019,damasceno_learning_2021}.

Motivated by these benefits, we introduce our vision of how family-based learning \cite{damasceno_learning_2019,damasceno_learning_2021} and testing \cite{classen_featured_2013,fragal_extending_2018} principles could be lifted to behavior-based fingerprint analysis. 
These notions should aid an efficient framework for family-based fingerprint analysis where a group of behavioral signatures are handled, matched and merged as a family model, rather than a group of individual signatures. 

\begin{example}{\textit{(Running example of behavioral variability models)}}
In Fig.~\ref{fig:ffsm}, we depict a family-based representation for the set of alternative product FSMs shown in the previous example. 

\begin{figure}
     \centering
     \includegraphics[width=\textwidth]{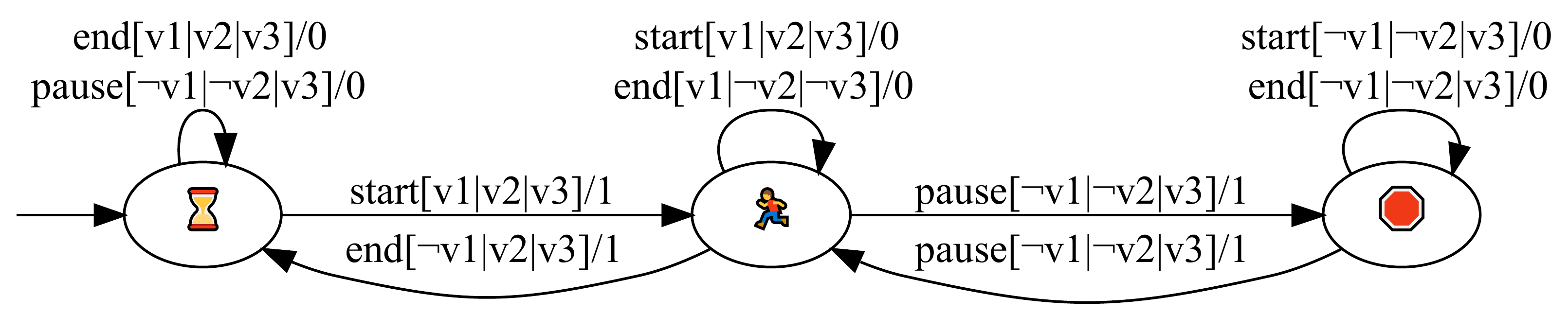}
     \caption{Example of family model expressed as a FFSM}
     \label{fig:ffsm}
\end{figure}
\end{example}

\subsection{A Framework for Family-Based Fingerprint Analysis}

\begin{figure}
    \centering
    \vspace{-0.3cm}
    \includegraphics[width=\linewidth]{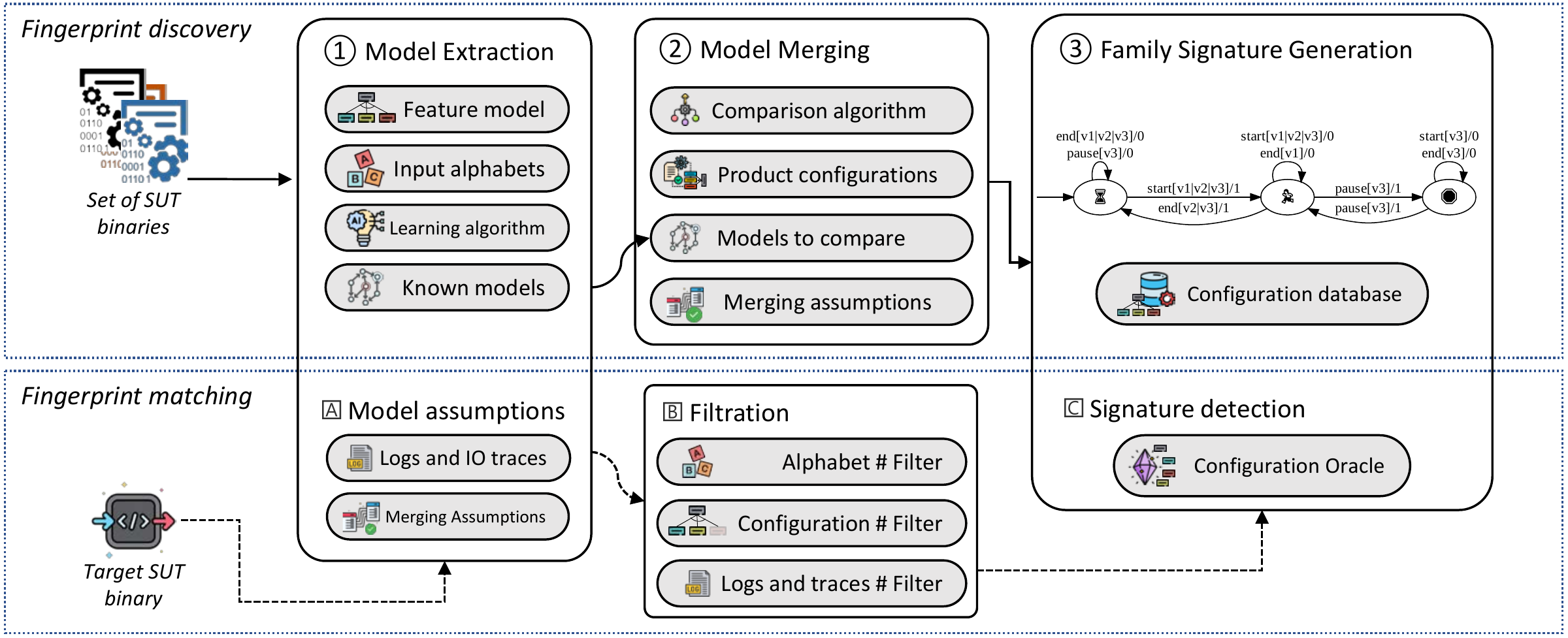}
    \vspace{-0.6cm}
    \caption{A framework for family-based fingerprint analysis}
    \vspace{-0.7cm}
    \label{fig:fb_fgpt:framework}
\end{figure}

In this paper, we propose the development of a framework for family-based fingerprint analysis.
We suggest principles from model learning \cite{damasceno_learning_2019,damasceno_learning_2021} and testing \cite{classen_featured_2013,fragal_extending_2018} as means to kick-off the automated creation and maintenance of family-based signatures from a set of SUT binaries. In Fig.~\ref{fig:fb_fgpt:framework}, we depict this framework, which, inspired by \cite{shu_formal_2011,shirani_binshape_2017}, we divided in two stages: 
(a) \textit{Fingerprint discovery}, where a family signature is generated by learning, matching, and merging SUT-specific signatures; and (b) \textit{Fingerprint Matching},  where the family signature is employed as a \textit{configuration oracle} to answer \textit{if} or \textit{under which circumstances} a given IO trace has been observed.

\subsubsection{Family Fingerprint Discovery}

When fingerprinting a set of SUT binaries that are akin, it is reasonable to assume that they share behavioral commonalities due to similar requirements or even reused components. Hence, we believe adaptive model learning \cite{huistra_adaptive_2018} is a variant that can aid in reducing the costs required for fingerprint discovery. In adaptive learning, pre-existing models are used to derive \texttt{MQ}s to steer learning algorithms to states maintained after updates, and potentially speed up the model learning process for systems evolving over time \cite{damasceno_learning_2019-1} and in space \cite{tavassoli2022:splc:adaptive_learning}. 
Hence, we believe these benefits may also hold in fingerprint discovery. 

Once a group of signatures is obtained, fingerprint matching may be performed in its standard way. However, as the cost for fingerprint group matching may increase exponentially to the number of alternative versions and the size of its candidate signatures, we suggest a model merging step to combine a set of behavioral signatures into a unified FFSM representation \cite{fragal_extending_2018}. To support this step, we find that state-based model comparison algorithms (e.g., LTS\_diff \cite{walkinshaw_automated_2013}, FFSM\_diff \cite{damasceno_learning_2021}) can provide efficient means to construct a \textit{family signature}. Merging assumptions can be used to preset state pairs matching \cite{walkinshaw_automated_2013} and aid the creation of a more succinct representation \cite{damasceno_learning_2021} for groups of fingerprints.
This concept of family signature provides the basis for a key entity in family-based fingerprinting experiments, namely the \textit{configuration oracle} (CQ). 

Our idea for a CQ is an abstract entity able to report \textit{if} or \textit{under which circumstances} (e.g., feature combinations, versions) a given IO trace has been previously observed. We believe that CQs can also be repurposed to recommend configurable test cases for distinguishing SUT versions from their \textit{observed outputs} or \textit{satisfiable presence conditions}. Thus, family-based signatures are amenable to be deployed in both passive and active fingerprint experiments for discovery and matching.

\subsubsection{Family Fingerprint Matching}

Once a family signature is created, variability-aware, model-based testing concepts can enable an efficient fingerprint matching. Particularly, we see that family-based testing principles, such as configurable test suites \cite{fragal_extending_2018}, could be repurposed as queries to check whether a particular IO trace has been previously observed. If so, the presence conditions assigned to the conditional transitions traversed by an IO trace can be used to constraint the configuration space of a family of SUT binaries, e.g., ``the following presence conditions must hold because the IO traces matches with this list of conditional state/transition''. 
To automate the task of fingerprint matching, SAT/SMT solvers can be used to reply what (or even how many) configurations can potentially match to a given SUT behavior, as \texttt{EQ}s do.

\begin{example}{\textit{(Example of fingerprint matching)}}
In Fig.~\ref{fig:conftest}, we illustrate an example of configurable test cases derived from the FFSM in Fig.~\ref{fig:ffsm}. 

\begin{figure}
     \centering
     \includegraphics[width=\textwidth]{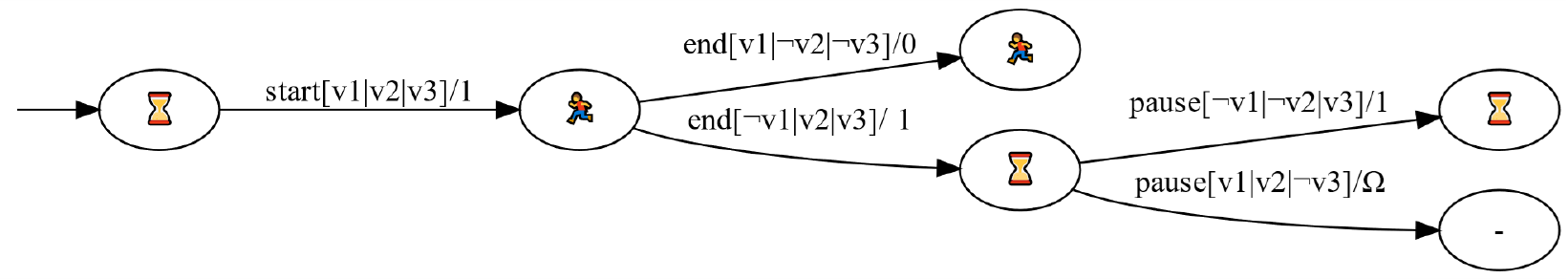}
     \caption{Example of configurable test case for fingerprint matching}
     \vspace{-0.3cm}
     \label{fig:conftest}
\end{figure}

From this configurable test case, we can find that the trace $\mathtt{start/1\cdot end/1}$ implies the constraint $ (v1|v2|v3) \land (\lnot v1|v2|v3)$ and, from it, we can discard a match between the SUT and version \texttt{v1}.
Also, we can find that this same input is able to distinguish versions \texttt{v1} and \texttt{v2}.
In this case, if the trace $\mathtt{start/1\cdot end/0}$ is observed, then the constraint $(v1|v2|v3) \land (v1 | \lnot v2| \lnot v3)$ is derived and hence, a match to \texttt{v1} is found.
\end{example}

\subsection{Practical and Theoretical Implications}

In this section, we outline a few implications of this framework on software analysis. 
These include 
(a) Combining passive and active fingerprinting experiments,
(b) Family-based fingerprinting in model learning, and
(c) Fingerprint Analysis in the Open-World. 

\subsubsection{Hybrid fingerprinting experiments.}
When fingerprinting, traces from passive experiments can be incorporated in fingerprint matching to constraint the configuration space of family-based fingerprints. Then, presence conditions derived from these IO traces can be used to steer fingerprint analysis to parts of the signature to reduce the uncertainty of what configuration is inside some unidentified SUT. Similar concepts have been used in adaptive learning to speed up update learning and should also aid performance improvements.

\subsubsection{Family Signatures In Active Model Learning.}
Family-based fingerprints may also support active model learning, particularly by providing \textit{EQ}s based on multiple merged hypotheses. Typically, equivalence queries are approximated via conformance testing techniques applied over a single hypothesis \cite{angluin_learning_1987}. However, some learning techniques may construct hypothesis non-deterministically \cite{vaandrager_new_2022} and hence, potentially lead to ``hypotheses mutants''. Aichernig et al. \cite{Aichernig2019} has shown that \texttt{EQ}s can be efficiently generated using mutation analysis over hypothesis. We believe these results may also hold when combined with family models. In fact, a similar idea has been already investigated by Devroey et al. \cite{devroey2016_fmbma} within the context of family model-based testing where behavioral variability models have been deployed to optimize the generation, configuration and execution of mutants. Nevertheless, there are still no studies deploying family model-based testing in active learning.

\subsubsection{Towards Fingerprinting Highly-Configurable Systems.} 
As our long term vision, we aim at making our approach suitable for highly-configurable systems, where it is infeasible to enumerate all variants or the complete SUT behavior. Hence, fingerprinting must rely on samples of traces. Currently, if the SUT does not have an \textit{exact match} with any signature, Shu et al. \cite{shu_formal_2011} recommends applying model learning \cite{angluin_learning_1987} to the SUT. However, in highly-configurable systems, exhaustive learning becomes impractical due to the potentially exponential number of valid configurations. Thus, it becomes interesting to inform whether an \textit{unindetified trace} has an \textit{inexact match} with patterns associated to a particular configuration or parameter. To address this, we believe that other variability-aware representations, e.g., composition-based models \cite{benduhn_survey_2015} or control-flow graphs \cite{rhein_variability-aware_2018}, and analysis techniques, e.g., statistical classification or clustering \cite{pereira_learning_2021}, may be more suitable to capture fingerprints as small behavioral or structural patterns, rather than an exact annotative-based model \cite{classen_featured_2013,fragal_validated_2017} of the SUT behavior. 

\section{Final Remarks}
\label{sec:concl}

This paper discusses a generic framework for lifting fingerprint analysis to the family-based level. We suggest that state-based model comparison algorithms \cite{walkinshaw_automated_2013} can aid the creation of concise FFSM representations \cite{damasceno_learning_2019,damasceno_learning_2021} from a set of fingerprints and enable efficient fingerprint analysis. We envision there are a plenty of real-world artifacts and alternative analysis and modeling approaches that could be used to start exploring and expanding this problem. Many artifacts are available in the Automata Wiki \cite{neider_benchmarks_2019}. We believe this repository constitutes a great opportunity to future investigations in this novel topic which we call \textbf{family-based fingerprinting analysis}.

\bibliographystyle{splncs04}
\bibliography{fvaan_fs60}

\end{document}